\documentclass[review]{elsarticle}

\usepackage{lineno,hyperref}
\modulolinenumbers[5]

\journal{Journal of \LaTeX\ Templates}

\begin{document}

\begin{frontmatter}

\title{Isoscalar and isovector pairing  in a formalism of quartets}

\author[address1]{M. Sambataro\corref{correspondingauthor}}
\cortext[correspondingauthor]{Corresponding author}
\ead{michelangelo.sambataro@ct.infn.it}

\author[address2]{N. Sandulescu}
\ead{sandulescu@theory.nipne.ro}

\author[address3]{C. W. Johnson}
\ead{cjohnson@mail.sdsu.edu}

\address[address1]{Istituto Nazionale di Fisica Nucleare - Sezione di Catania, Via S. Sofia 64, I-95123 Catania, Italy}
\address[address2]{National Institute of Physics and Nuclear Engineering, P.O. Box MG-6, Magurele, Bucharest, Romania}
\address[address3]{Department of Physics, San Diego State University, 5500 Campanile Drive, San Diego, California 92182-1233, USA}

\begin{abstract}
Isoscalar $(T=0,J=1)$ and isovector $(T=1,J=0)$ pairing  correlations in the ground state of self-conjugate nuclei are treated in terms of alpha-like quartets  built by two protons and two neutrons coupled to total  isospin $T=0$ and total angular momentum $J=0$. 
Quartets are constructed dynamically via an iterative variational procedure and the ground state is represented as a product of such quartets. 
It is shown that the quartet formalism
%ns
describes accurately the ground state
%ns
energies of realistic isovector plus isoscalar pairing Hamiltonians in nuclei 
with valence particles outside the $^{16}$O, $^{40}$Ca and $^{100}$Sn cores. 
%ni
Within the quartet formalism we analyse  the competition 
between isovector and isoscalar pairing correlations and find that for nuclei
with the valence nucleons above the cores $^{40}$Ca and $^{100}$Sn the isovector 
correlations account for the largest fraction of the total
pairing correlations. This is not the case for $sd$-shell nuclei for which 
isoscalar correlations  prevail. 
Contrary to many mean-field studies, isovector and isoscalar pairing correlations  mix significantly in the quartet approach.
\end{abstract}


\end{frontmatter}

%\linenumbers

\section{Introduction}
One of the most debated and yet open issues in nuclear physics is whether or not
the  deuteron-like proton-neutron pairs of isospin $T=0$ and angular momentum $J=1$
behave coherently in the form of a condensate, analogous in structure to the
condensates of like-particle pairs. For about 50 years the isoscalar proton-neutron 
pairing and its competition with the isovector $(T=1,J=0)$  pairing have been commonly studied
in the framework of the Hartree-Fock-Bogoliubov (HFB)  theory. Most of its developments, starting from the
pioneering works \cite{goswami_kisslinger,chen_goswami, goodman_struble,
wolter} have been reviewed by Goodman \cite{goodman_adv,goodman1990}  
(for a recent study, see  \cite{gezerlis}). However, as clearly evidenced in applications within exactly solvable models of $T=1$ and $T=0$ pairing \cite{dobes,evans,engel,lerma}, this theory suffers important limitations due to its inherent violations of the particle number and of the isospin. Such violations are, of course, absent in the Shell Model (SM) and various attempts have been made to employ this approach to elucidate the 
competition between the  isoscalar and isovector pairing correlations \cite{poves,pittel_sandulescu}. However, it is still unclear how 
one could identify in the complicated SM wave function the existence of the collective pairs 
and their possible coherence in the form of a pair condensate. 

In the present study we propose  a new approach for treating the isoscalar and isovector
pairing interaction in $N=Z$ nuclei which is  based not on pairs, as in the case of the HFB theory, 
but on alpha-like quartets. This approach presents the advantage  of conserving exactly the 
particle number and the isospin and, at the same time, it is  simple enough  for understanding
the role played by the  isoscalar and isovector pairing correlations. 

The idea of using quartets for describing  proton-neutron pairing in nuclei is rather old \cite{lane} but it
has been mostly employed for treating the isovector interaction
 \cite{flowers,zelevinsky,yamamura,talmi,hasegawa}.    A consistent quartet formalism for treating the isovector pairing, which 
 conserves the particle number, the isospin and takes into account exactly the 
 Pauli blocking, has been  proposed in Refs. \cite{qcm1,qcm2} . In this model the ground state
  of $N=Z$ nuclei is approximated by a condensate of alpha-type quartets formed by
  two isovector pairs coupled to $T=0$.  Recently this model  has been generalized by allowing
  the isovector quartets to be different from one another \cite{sasa_t1}.
  In the present letter we extend the quartet model
  %ns\cite{sasa_t1}
  to the treatment of both the  isovector and  the isoscalar pairing interactions in
  nuclei with  an equal number  of protons and neutrons outside a self-conjugate core. 
   
The manuscript is structured as follows. In Section 2, the ground state of the isovector plus isoscalar Hamiltonian for $N=Z$ systems is formulated in the formalism of quartets.
In Section 3,
%ns
the quartet formalism is applied to nuclei with valence particles outside the  $^{16}$O, $^{40}$Ca and $^{100}$Sn. 
Finally, in Section 4, we give the  conclusions.
   
\section{The quartet formalism}
   The  isovector plus isoscalar pairing Hamiltonian in a spherically symmetric mean field has the form
\begin{equation}
H=\sum_i  \epsilon_i N_i + 
\sum_{i,j} V^{T=1}_{J=0} (i,j) \sum_{T_z}P^+_{i,T_z} P_{j,T_z}+
\sum_{i\leq j,k\leq l} V^{T=0}_{J=1}(ij,kl) 
\sum_{J_z}D^+_{ij,J_z} D_{kl,J_z}.
\end{equation}
In the first term, $\epsilon_i$ and $N_i$ are, respectively, the energy and the particle number operator relative to the single-particle state $i$. The symbol $i$ is a short cut notation for
$\{n_i,l_i,j_i,\tau_i\}$, with $\{n_i,l_i,j_i\}$ being the standard orbital quantum numbers and $\tau_i$ denoting the isospin projection. The Coulomb interaction
between the protons is not taken into account so that the single-particle energies of  protons 
and neutrons are assumed to be equal.
 The second term in Eq. (1) is the isovector pairing interaction. This is formulated
in terms of the  non-collective pair operators
\begin{equation}
P^+_{i,T_z}= \frac{1}{\sqrt{2}}[a^+_i a^+_i ]^{T=1,J=0}_{T_z}
\end{equation}
where $T_z$ denotes  the three projections of the isospin $T=1$ corresponding to
neutron-neutron ($T_z=1$), proton-proton ($T_z=-1$) and proton-neutron ($T_z=0$) pairs.
The isoscalar pairing interaction, the third term in Eq. (1),  is written in terms of the pair operators
\begin{equation}
D^+_{ij,J_z}=\frac{1}{\sqrt{1+\delta_{ij}}}[a^+_i a^+_j ]^{J=1,T=0}_{J_z}
\end{equation}
where  $J_z$ denotes the three projections of the angular momentum $J=1$.

It is worth mentioning that the Hamiltonian (1) is exactly solvable only for
%ns
$V^{T=1}_{J=0}(i,j)= V^{T=0}_{J=1}(ij,kl)=g$, where $g$ is a state-independent 
pairing strength, and in the absence of the spin-orbit interaction.
In this case, the isovector  and isoscalar correlations play a similar role and 
contribute to the ground  state energy to an equal amount \cite{lerma}.

In this work we investigate to which extent the ground state of the Hamiltonian (1) for an even-even self-conjugate nucleus 
can be represented in terms collective alpha-like quartets having total angular momentum $J=0$ and total isospin $T=0$. One can form two types of quartets: isovector quartets, resulting from the coupling of two isovector pairs (2),
\begin{equation}
Q^{+ (iv)}_{\nu} = \sum_{i,j} x^{(\nu )}_{ij} [P^+_i P^+_j]^{T=0}
\end{equation}
 and isoscalar quartets, formed instead by two isoscalar 
 pairs (3)
\begin{equation}
Q^{+ (is)}_{\nu} = \sum_{ij,kl} y^{(\nu )}_{ij,kl} [D^+_{ij} D^+_{kl}]^{J=0}.
\end{equation}
By summing up these quartets one constructs the generalized quartets
\begin{equation}
Q^+_{\nu}=Q^{+ (iv)}_{\nu} + Q^{+(is)}_{\nu}. 
\end{equation}
We approximate the ground
state of the Hamiltonian (1) for an even-even $N=Z$ nucleus as a product of such quartets, namely
\begin{equation}
|\Psi_{gs}\rangle \equiv |QM \rangle =\prod^{N_Q}_{\nu =1}Q^\dag_\nu |0\rangle 
\end{equation}
%ns
where $|0\rangle$ denotes a self-conjugate core of nucleons not affected by the pairing interaction.
Since each quartet has $T=0$ and $J=0$, these also represent the quantum numbers of the ground state (7).
%ns
Dealing with $T=0$ and $J=0$ quartets only has the great advantage of 
not requiring any angular momentum coupling, but also is simpler to apply than that proposed in Ref. \cite{arima_gillet}
which instead employed general quartets with $J \neq 0$, $T \neq 0$.

The QM state  depends on the mixing amplitudes $x^{(\nu)}_{ii'}$ and $y^{(\nu)}_{ii',jj'}$
 which define the collective isovector and isoscalar quartets.  In order to find them 
 we employ a generalization of the iterative variational procedure used in the case of the
 isovector pairing \cite{sasa_t1} (for details, see also \cite{sasa,samba2}). The procedure 
consists of a sequence of diagonalizations of the Hamiltonian (1) in 
spaces whose size $N_q$ is given by the total number of non-collective isovector 
($[P^+_i P^+_j]^{T=0}$) and
isoscalar ($[D^+_{ij} D^+_{kl}]^{J=0}$) quartets which can be formed in the chosen model space of single-particle
states. For simplicity, we denote all these non-collective quartets as $q^+_{\mu}$
($\mu=1,2 ..{\it N_q}$) and write the collective quartet (6) generically as $Q^+_\nu=\sum_\mu c^{(\nu)}_\mu q^+_{\mu}$.  In order to describe a system with $N_Q$ quartets, we
proceed step-by-step starting from the case of  one quartet. For $N_Q=1$,   the 
Hamiltonian (1) is diagonalized in the space $F_1$ spanned by  all possible non-collective quartets, i.e.
$
F_1= \Bigl\{ q^+_\mu  |0\rangle \Bigr\}. 
$ 
The lowest state in energy which results from this diagonalization
represents the exact ground state for the system with two neutrons and two protons and it has the form
$
|\Psi_1\rangle =Q^+_1 |0\rangle .
$
For the system with $N_Q=2$ quartets, as a first approximation of the ground state, we assume the lowest state in energy resulting from the diagonalization of $H$ in the space
 $
 F^{(1)}_2=\Bigl\{q_\mu  Q^+_1 |0\rangle \Bigr\} ,
$
 where $Q^+_1$ is the quartet previously determined.
 This state has therefore the form
 \begin{equation}
 |\Psi^{(1)}_2\rangle = Q^+_2 Q^+_1 |0\rangle \equiv Q^+_2 |\Psi_1\rangle .
 \end{equation}
From this point on, a series of diagonalizations starts whose purpose is that of finding the quartets which guarantee the lowest possible energy of the state (8). Each diagonalization is meant to update one quartet while leaving the other unchanged. In the second step, for instance, one proceeds by diagonalizing $H$ in the space
$F^{(2)}_2=\Bigr\{q^+_\mu Q^+_2 |0>\Bigr\}$. This diagonalization 
generates the second order approximation for the ground state
 \begin{equation}
 |\Psi^{(2)}_2\rangle = Q^{+(new)}_1 Q^+_2 |0\rangle .
 \end{equation} 
This state is expected to be lower (or, at worst, equal) in energy with respect to $|\Psi^{(1)}_2\rangle$ and so each diagonalization, while updating a quartet, drives the two-quartet state toward its minimum in energy. This diagonalization is iterated until the energy converges. 

The procedure  illustrated  in the previous paragraph for the case $N_Q =2$ can be generalized for any value of $N_Q$.
In general, if $Q^\dag_\nu$ $(\nu =1,2,...,N_Q-1)$ are the final quartets generated for the system $N_{Q}-1$, 
we start by finding
the lowest order approximation of the ground state for the system with $N_Q$ quartets, which results from the diagonalization of $H$ in the space
 \begin{equation}
F^{(1)}_{N_Q}=\Biggl\{ q^+_\mu
\prod^{N_Q-1}_{\nu=1}Q^\dag_\nu |0\rangle 
\Biggr\}
\equiv \Biggl\{ q^+_\mu |\Psi_{N_Q-1}\rangle\Biggr\}
\end{equation}
and it has therefore the form
 \begin{equation}
 |\Psi^{(1)}_{N_Q}\rangle =Q^+_{N_Q} |\Psi_{N_Q-1}\rangle .
 \end{equation}
This lowest order approximation is improved by an iterative sequence of diagonalizations which updates the quartets one by one and drives this state towards its minimum in energy. 

It is worthy noticing that, owing to this continuous updating, the quartets that populate the final state $|\Psi_{N_Q}\rangle$ are different from those defining $|\Psi_{N_Q-1}\rangle$. 
In this sense, quartets are generated dynamically for every $N_Q$.
This fact makes impossible to establish a simple connection between $|\Psi_{N_Q}\rangle$ and 
$|\Psi_{N_Q-1}\rangle$. However, as  evidenced in Eq. (11), if the iterative procedure is arrested at the lowest order, the ground state at this stage, $|\Psi^{(1)}_{N_Q}\rangle$,
simply results from the action of a quartet creation operator on the ground state for $N_Q-1$. This expression is of particular interest because, provided that $|\Psi^{(1)}_{N_Q}\rangle$ 
can be proved to be a good approximation of the exact ground state,
it would give a clear evidence of the key role played by $T=0$, $J=0$ quartets in the ground state of the isovector-isoscalar pairing Hamiltonian.

\section{Results}
To test the accuracy of the quartet model  we have performed calculations 
for three sets of $N=Z$ nuclei with valence nucleons outside the 
 $^{16}$O, $^{40}$Ca, and $^{100}$Sn cores.  The isovector and isoscalar 
 pairing forces of the Hamiltonian (1) have been extracted, respectively, from the $(T = 1, J = 0)$  and $(T=0,J=1)$ components of standard shell model interactions. More precisely, 
for nuclei outside the $^{16}$O core we have used the  
USDB interaction  \cite{usd},  for those outside the $^{40}$Ca core the monopole-modified Kuo-Brown 
 interaction KB3G \cite{poves} and, for those outside the $^{100}$Sn core, the 
effective G-matrix interaction of Ref. \cite{gmatrix}.
As single-particle energies we have taken those employed with the previous
interactions (e.g., see  Ref. \cite{qcm1}).

The results for the pairing correlation energy, defined as the difference
between the ground state energies obtained with and without the
pairing force, are given in Table 1. In order to check the accuracy of the
quartet model, the calculations have been done only for those  $N=Z$ nuclei for
which the Hamiltonian (1) could be diagonalized exactly.
As seen in Table 1, the errors relative to the exact solution are very small,
under $1\%$. This shows that the ansatz (7) for the ground state is a very
good approximation for describing the isoscalar-isovector pairing correlations.

In Table 1 we also present the results relative to the lowest order approximation (11). 
In addition to the correlation energies, we show the overlaps between 
this approximated ground state and the actual QM ground state, i.e. the state at the end of the iterative process. It can be seen that the relative errors remain confined within $1\%$
even in this case and that these overlaps are very close to 1. Therefore the lowest order approximation too emerges as an excellent approximation of the exact ground state. 
  
\begin{table}[hbt]
\caption{Ground state correlation energies (in MeV) calculated for the isovector plus isoscalar
pairing Hamiltonian (1) with strengths extracted from standard shell
model interactions (see text). The results are shown for the exact  diagonalization, 
the QM  state (7) and the lowest order approximation (11),  denoted by QM(l.o.). 
The errors relative to the exact results are given in brackets.
Overlaps (in absolute values) between 
the states QM and QM(l.o.) are reported in the last column.}
\begin{center}
\begin{tabular}{|c|c|c|c|c|}
\hline
\hline
   &    Exact & QM & QM(l.o.) & $\langle QM|QM(l.o.)\rangle$ \\
\hline
\hline
$^{24}$Mg  &  28.694  & 28.626 (0.24$\%$)   & 28.592 (0.35$\%$)  & 0.9993 \\
$^{28}$Si  &  35.600  & 35.396 (0.57$\%$)  & 35.307 (0.82$\%$)   & 0.9980  \\
$^{32}$S   &  38.965 & 38.865 (0.25$\%$)  & 38.668 (0.76$\%$)  &  0.9942  \\
\hline
$^{48}$Cr &  11.649   & 11.624 (0.21$\%$)   & 11.614 (0.30$\%$)   &  0.9996 \\
$^{52}$Fe &  13.887   & 13.828 (0.43$\%$)  & 13.804 (0.60$\%$)   &  0.9994 \\
\hline
$^{108}$Xe &  5.505   & 5.495 (0.18$\%$)     & 5.490 (0.27$\%$)   & 0.9995  \\
$^{112}$Ba &  7.059    & 7.035 (0.34$\%$)          & 7.025 (0.48$\%$)  & 0.9987 \\

\hline
\hline
\end{tabular}
\end{center}
\end{table}

Having verified that the QM state (7) is able to describe with very high precision the pairing correlation energies of the isovector plus isoscalar Hamiltonian (1), the quartet formalism can be used to analyse the competition between the isovector and isoscalar components of the pairing interaction. Due to the mixed nature of the quartets, Eq. (6), the QM ground state (7) contains an isovector component
 \begin{equation}
|iv\rangle =\prod^{N_Q}_{\nu =1}Q^{\dag (iv)}_{\nu} |0\rangle ,
\end{equation}
an isoscalar component
\begin{equation}
|is\rangle =\prod^{N_Q}_{\nu =1}Q^{\dag (is)}_{\nu} |0\rangle 
\end{equation}
and,  for $N_Q >1$, a mixed component with both isovector and isoscalar quartets.
As all these components are not orthogonal to each other, it is not trivial to analyse their competition in the ground state.  Thus in order to explore the relative importance of the isovector and isoscalar correlations, we have carried out two further QM calculations, one by assuming a ground state formed only by isovector quartets, i.e. of the type (12), and the other with a ground state formed only by isoscalar quartets, i.e. of the type (13). 
The results of these calculations are presented in Table 2 where we report the ground state correlation  energies in the different approximations and the overlaps  between the corresponding wave functions. One can see that, for nuclei with valence nucleons outside the  $^{40}$Ca and $^{100}$Sn cores, the isovector quartet state (12) is able to account for the largest part of the correlation energy induced by the isovector-isoscalar interaction.
 This fact is also supported by the large overlaps  with the QM state (7). However, the isoscalar correlation contribution remains non-negligible because, as seen in Table 1, it reduces the errors in the correlation energies by about one order of magnitude. 
A different situation is observed instead in the case of $sd$-nuclei where the pairing forces extracted from the USDB shell-model interaction give rise to a prominence of the isoscalar contribution. Still in Table 2 one can notice that the overlap between the isovector-type (12) and isocalar-type (13) ground states can be rather large. This overlap is a measure of the difficulty in disentangling the isovector and isoscalar contributions. It is worth mentioning that in the present symmetry conserving quartet
formalism the isovector and isoscalar pairing correlations always coexist, which is  usually not the case in HFB calculations \cite{goodman1990,gezerlis}.    
    
\begin{table}[hbt]
\caption{ Correlation energies (in MeV) calculated with the  isovector quartet state (12) and the
isoscalar quartet state (13). In the first column we give, as a reference, the results
corresponding to the full QM state (7).  The errors relative to the QM
 results are shown in brackets. In the three columns on the right we report the overlaps (in absolute values) between the quartet states just mentioned.}
 \begin{center}
\begin{tabular}{|c|c|c|c|c|c|c|}
\hline
\hline
 &  QM & iv & is &$ \langle QM|iv\rangle$ & $\langle QM|is\rangle$ & $\langle iv|is\rangle$ \\
\hline
\hline
$^{20}$Ne &  15.985      &  14.402 (9.9$\%$)  &   15.130 (5.4$\%$)   &     0.884   &    0.953     &  0.843  \\ 
$^{24}$Mg &  28.626      &   23.269 (18.7$\%$) &   26.925 (5.9$\%$)   &     0.650   &    0.911 &  0.336   \\ 
$^{28}$Si &  35.396        &   28.897 (18.4$\%$) &   33.376 (5.7$\%$)  &     0.590    &   0.911     &  0.343  \\ 
$^{32}$S  &  38.865        &   33.959 (12.6$\%$) &   37.884 (2.5$\%$)   &     0.638    &   0.973     &  0.595  \\
\hline
$^{44}$Ti &  7.019   &  6.274 (10.6$\%$)   & 4.917 (30.0$\%$) & 0.901 & 0.678 & 0.303  \\ 
$^{48}$Cr &  11.624   & 10.589 (8.9$\%$)   & 7.384 (36.5$\%$) & 0.906 & 0.497 & 0.221   \\ 
$^{52}$Fe &  13.828   & 12.814 (7.3$\%$)  & 9.980 (27.8$\%$) & 0.927 & 0.753 & 0.746  \\ 
\hline
$^{104}$Te &      3.147  & 3.041 (3.4$\%$)   &  1.549 (50.8$\%$) & 0.978    &  0.489     &  0.314   \\ 
$^{108}$Xe &  5.495   &   5.240  (4.6$\%$) &  2.627 (52.2$\%$) &  0.958   &  0.354     &  0.234   \\ 
$^{112}$Ba &  7.035    &   6.614  (6.0$\%$) & 4.466 (36.5$\%$)   & 0.939   &  0.375    &  0.376   \\
\hline
\hline
\end{tabular}
\end{center}
\end{table}

Previous works (e.g., see Refs. \cite{gezerlis,pittel_sandulescu}) have evidenced a strong effect 
of the spin-orbit interaction on the interplay between isovector and isoscalar correlations.
We have investigated this effect in the case of $pf$-shell nuclei by repeating 
the QM calculations in the absence of the single-particle energy splittings induced
by the spin-orbit interaction. In particular, we have assumed all single particle energies equal to 2.6 MeV (roughly speaking the centroid of the original single particle energies \cite{poves}) and kept unchanged the isovector and isoscalar strengths in the Hamiltonian (1). The new results appear to be reversed with respect to those shown in Table 2, with the isoscalar quartet state (13) accounting for the largest fraction of the correlation energy induced by the isovector-isoscalar interaction (the deviations from the QM values are now confined within 5$\%$ while they become larger than 20$\%$ for the isovector state (12)). Also the overlaps between the corresponding states appear to be reversed with $\langle QM|is\rangle$ being now close to 0.9. Our analysis within the quartet formalism therefore confirms that isoscalar correlations are strongly hindered by the spin-orbit interaction in these nuclei.

\section{Conclusions}
In this work we have described the ground state of the isovector plus isoscalar pairing Hamiltonian in even-even $N=Z$ nuclei in a formalism of alpha-like quartets. Quartets are built by two neutrons and two protons coupled to total isospin $T=0$ and total angular momentum $J=0$. The ground state is represented as a product of quartets and a
procedure to construct them has been described. The formalism does not violate any symmetry of the Hamiltonian. We have carried out a number of numerical tests for systems
with valence nucleons outside the $^{16}$O, $^{40}$Ca and $^{100}$Sn cores and with pairing interactions extracted from realistic shell model Hamiltonians.
We have verified that ground state correlation energies are reproduced with high accuracy 
in the quartet formalism. 
%ns
For the same systems we have  shown that, to a very good extent,
the $T=0$, $J=0$ quartets link the pairing ground states of adjacent even-even $N=Z$ nuclei. 
Therefore the role played by these quartets in even-even self-conjugate nuclei appears 
analogous to that of Cooper pairs in the ground state of a like-particle pairing Hamiltonian. 
We have also analyzed the competition between the isovector and isoscalar pairing within the quartet formalism. Isovector pairing
correlations have been found dominant in the ground states of $pf$-shell nuclei and of nuclei outside the $^{100}$Sn core while, in $sd$-shell nuclei, the isoscalar pairing correlations have been found to prevail. A strong mixing between isovector and isoscalar pairing correlations has been observed in most of the cases. Finally, we have analyzed the effect of the spin-orbit interaction on the interplay between isovector and isoscalar correlations in $pf$-shell nuclei. Consistently with previous works, we have found that this interplay is strongly affected by this interaction and that, in his absence, isoscalar correlations become the dominant ones.

\vskip 0.3cm

{\it Acknowledgments} 
This work was supported by the Romanian Ministry of Education and Research
through the grant Idei nr 57 and by the U.S. Department of Energy, Office of Science, Office of Nuclear Physics, 
under Award Number  DE-FG02-96ER40985.

\section*{References}

\bibliography{biblio}

\end{document}